\documentclass[a4paper, 12pt]{article}

\usepackage[sort&compress]{natbib}
\bibpunct{(}{)}{;}{a}{}{,} 

\usepackage{amsthm, amsmath, amssymb, mathrsfs, multirow, url, subfigure}
\usepackage{graphicx} 
\usepackage{ifthen} 
\usepackage{amsfonts}
\usepackage[usenames]{color}
\usepackage{fullpage}

\theoremstyle{plain}

\theoremstyle{definition}

\theoremstyle{remark}

\newcommand{\prob}{\mathsf{P}}

\renewcommand{\phi}{\varphi}

\newcommand{\nm}{\mathsf{N}}
\newcommand{\bet}{\mathsf{Beta}}

\newcommand{\bin}{\mathsf{Bin}}

\newcommand{\chisq}{\mathsf{ChiSq}}
\newcommand{\unif}{\mathsf{Unif}}
\newcommand{\gam}{\mathsf{Gamma}}

\newcommand{\Ybar}{\bar{Y}}
\newcommand{\ybar}{\bar y}

\title{A statistical inference course based on p-values}
\author{
Ryan Martin \\
Department of Mathematics, Statistics, and Computer Science \\
University of Illinois at Chicago \\
\url{rgmartin@uic.edu} 
}
\date{\today}

\begin{document}

\maketitle 


\begin{abstract}
Introductory statistical inference texts and courses treat the point estimation, hypothesis testing, and interval estimation problems separately, with primary emphasis on large-sample approximations.  Here I present an alternative approach to teaching this course, built around p-values, emphasizing provably valid inference for all sample sizes.  Details about computation and marginalization are also provided, with several illustrative examples, along with a course outline.  

\smallskip

\end{abstract}

\section{Introduction}
\label{S:intro}

Consider a first course in statistical inference, whose target audience is upper-level undergraduate and beginning graduate students in statistics or other quantitative fields such as computer science, economics, engineering, and finance.  These students typically have been exposed to some basic statistical methods, such as t-tests, in a previous course.  Moreover, the course is question is usually the second course in a ``probability + statistics'' sequence, so the students are assumed to have background in (calculus-based) probability, which covers random variables and their distributional properties; in particular, students will have had at least an introduction to sampling distributions and key results like the law of large numbers and central limit theorem.  Commonly used textbooks for this statistics theory course include: \citet{casella.berger.book}, \citet{wackerly.book}, and \citet{hogg.mckean.craig.book}.  A typical course outline for the first statistical theory course starts with a review of sampling distributions and proceeds with details about point estimation, hypothesis testing, and confidence intervals, in turn.  As a result of this structure, and the amount of material to be covered, the majority of the course is spent on point estimation and its properties, e.g., unbiasedness, consistency, etc, leaving very little time at the end to cover hypothesis testing and confidence intervals.  This is unfortunate because it gives students the wrong impression that point estimation is the priority, with hypothesis tests and confidence intervals of only secondary importance.  For statistics students, this skewed perspective would eventually get straightened out in their more advanced courses.  For the non-statistics students, however, the course in question may be their only serious exposure to statistical theory, so it is essential that it be more efficient, focusing more on the essentials of statistical inference, rather than unnecessary and out-dated technical details.  In this paper, I will describe a new approach to presenting the core concepts of a statistical inference course, built around the familiar p-value, and inspired in part by recent work in \citet{imbook}.  

Despite the controversies surrounding hypothesis testing and p-values \citep[e.g.,][]{fidler.etal.2004, schervish1996}, including the recent ban of p-values in \emph{Basic and Applied Social Psychology} \citep{pvalue.ban}, statisticians know the value of these tools; see the official statement\footnote{\url{http://amstat.tandfonline.com/doi/abs/10.1080/00031305.2016.1154108}} from the American Statistical Association, along with comments.  In particular, p-values can be used to construct hypothesis tests with desired frequentist Type~I error rate control, and these p-value-based tests can be inverted to obtain a corresponding confidence interval.  It is in this sense that my proposed framework is built around the p-value so, from a technical point of view, there is nothing new or surprising presented here.  However, there are a number of important consequences of the proposed approach.  
\begin{itemize}
\item P-values are familiar to students from their basic statistics course(s), so the transition to using p-values in a more fundamental way in this course ought to be relatively smooth.  Students understanding what a p-value \emph{means} is not essential.

\vspace{-2mm}

\item Hypothesis testing, confidence intervals, and point estimation (if necessary) can all be handled via a single p-value function which streamlines the presentation.    
 
\vspace{-2mm}

\item Core topics in a statistics theory course, such as likelihood, maximum likelihood, and sufficiency, fit naturally in the proposed course through the construction and computation of the p-value function.  

\vspace{-2mm}

\item For simple examples, the p-value function can be computed analytically, and this allows the instructor to cover the standard distribution theory results, in particular, pivots.  Beyond the simple examples, numerical methods are needed, and this provides an opportunity for statistical software \citep[e.g., R,][]{Rmanual} and Monte Carlo methods to be presented and applied in a statistics theory course. 
 
\vspace{-2mm}

\item All the usual examples can be solved either analytically or numerically, so asymptotic theory would not be a high priority in the proposed course.  Indeed, the role of asymptotic theory is just to demonstrate a unification of the examples and to provide simple p-value function approximations.  
\end{itemize}
Overall, I believe that this p-value-centric course, which puts primary focus on provably valid inference and computation, strikes the right balance between what is covered in a standard statistics theory course and the modern ideas and tools that students need.  Moreover, it provides a more accurate picture of what statistical inference is about, compared to the traditional message that over-emphasizes asymptotic approximations.  

The remainder of the paper is organized as follows.  Section~\ref{S:inference} sets the notation, gives some background on the p-value, and presents the basic p-value-based approach.  The key is that it is conceptually straightforward to construct tests and confidence regions based on the p-value that are provably exact, or at least conservative.  This p-value function can be computed analytically in only a few textbook examples (see Section~\ref{SS:examples1}), so more sophisticated tools are needed.  Section~\ref{S:beyond} presents some details that go beyond the basics, including Monte Carlo methods for evaluating the p-value function, techniques for handling nuisance parameters, and asymptotic approximations.  Some more challenging examples are presented in Section~\ref{S:examples2}, and Section~\ref{S:outline} provides a sketch of a course outline.  Some concluding remarks are given in Section~\ref{S:discuss}.  R code for the examples is available as {Supplementary Material}.


\section{Inference based on p-values---the basics}
\label{S:inference}

\subsection{Key ideas}
\label{SS:main}

Suppose we have observable data $Y$ with sampling model $\prob_\theta$, known up to the value of the parameter $\theta$, which takes values in $\Theta$.  Both the data and parameter can be vectors, and it is not necessary to assume independence, etc.  Arguably the most fundamental statistical problem is hypothesis testing, and the simplest version takes the null hypothesis as $H_0: \theta = \theta_0$, for fixed $\theta_0 \in \Theta$; for the alternative hypothesis, here I take $H_1: \theta \neq \theta_0$, but other choices can be made depending on the context.  In my experience, students can relate to this problem and the logic behind the solution, i.e., $H_0$ identifies our ``expectations,'' and if the observation differs too much from these expectations, then there is doubt about the truthfulness of $H_0$.  More formally, consider a test statistic $T_{\theta_0}(Y)$ and, without loss of generality, assume that large values of $T_{\theta_0}(Y)$ cast doubt on $H_0$, suggesting that $H_0$ be rejected.  As a measure of the amount of support in observed data $Y=y$ in the truthfulness of $H_0: \theta = \theta_0$, consider the p-value
\begin{equation}
\label{eq:pvalue}
p_y(\theta_0) = \prob_{\theta_0}\{T_{\theta_0}(Y) \geq T_{\theta_0}(y) \}.
\end{equation}
It is well-known that the p-value is \emph{not} the probability that $H_0$ is true, but it does carry some relevant information, i.e., if $p_y(\theta_0)$ is small, then the observation $y$ is extreme compared to expectations under $H_0$, thereby casting doubt on $H_0$.  This intuition can be used to develop a formal testing rule, that is, one can reject $H_0$, based on observation $Y=y$, if and only if $p_y(\theta_0) \leq \alpha$, where $\alpha \in (0,1)$ is a pre-determined significance level.  It can easily be shown that this rule has Type~I error probability $\leq \alpha$; if the null distribution of $T_{\theta_0}(Y)$ is continuous, then equality is attained.  

I have focused so far on simple null hypotheses, i.e., $H_0: \theta = \theta_0$, but more general cases can be handled similarly.  Indeed, if the null hypothesis is $H_0: \theta \in \Theta_0$, for some subset $\Theta_0$ of $\Theta$, then, with a slight abuse of notation, the p-value is expressed as 
\begin{equation}
\label{eq:pvalue.sup}
p_y(\Theta_0) = \sup_{\vartheta \in \Theta_0} p_y(\vartheta), 
\end{equation}
the largest of the p-values associated with a simple null consistent with $\Theta_0$.  

The jumping off point here is that the p-value does not need to be tied to a specific null hypothesis.  That is, define $T_\theta(Y)$ as a function of data $Y$ and parameter $\theta$ and define the \emph{p-value function} 
\begin{equation}
\label{eq:function}
p_y(\theta) = \prob_{\theta}\{T_{\theta}(Y) \geq T_{\theta}(y) \}, \quad \theta \in \Theta. 
\end{equation}
In other contexts, the p-value function has been given a different name, e.g., \emph{preference functions} \citep{spjotvoll1983}, \emph{confidence curves} \citep{blaker.spjotvoll.2000, schweder.hjort.2002, schweder.hjort.book, xie.singh.2012, birnbaum1961}, \emph{significance functions} \citep{fraser1991}, and \emph{plausibility functions} \citep{plausfn}.  I prefer the latter name because it has a nice interpretation, though here I stick with ``p-value function'' because that name is familiar to students and is commonly used in the literature.  Names aside, the key observation is that the distributional properties of the p-value used above to justify the performance of the test extend in a natural way beyond the hypothesis testing context.  That is, 
\begin{equation}
\label{eq:valid}
\prob_\theta\{p_Y(\theta) \leq \alpha\} \leq \alpha, \quad \alpha \in (0,1), \quad \theta \in \Theta.
\end{equation}
As a particular application, take a fixed $\alpha \in (0,1)$ and define the set 
\begin{equation}
\label{eq:region}
C_\alpha(y) = \{\theta: p_y(\theta) > \alpha\}. 
\end{equation}
This can be interpreted as the set of all $\theta$ values which are ``sufficiently plausible'' given the observation $Y=y$.  Formally, it follows from \eqref{eq:valid} that $C_\alpha(Y)$ is a $100(1-\alpha)$\% confidence region for $\theta$ in the sense that the coverage probability is at least $1-\alpha$; coverage is exact if $T_\theta(Y)$ has a continuous distribution.  More abstractly, for any $A \subset \Theta$, one can view $p_y(A)$, defined as in \eqref{eq:pvalue.sup}, as a measure of how plausible is the claim ``$\theta \in A$'' based on observation $Y=y$, and the distributional results above guarantee a particular validity or calibration property: in standard terms, a test which rejects $H_0: \theta \in A$ when $p_y(A) \leq \alpha$ will control Type~I error at level $\alpha$.  

If desired, one can also construct a point estimator based on the p-value function by solving the equation $p_y(\theta) = 1$ for $\theta$.  In terms of the confidence region in \eqref{eq:region}, this value of $\theta$ is one that is contained in \emph{all} $100(1-\alpha)$\% confidence regions as $\alpha$ ranges over $(0,1)$.  As an example, suppose that $T_\theta(y)$ is the likelihood ratio statistic, defined as 
\begin{equation}
\label{eq:lrt}
T_{\theta}(y) = \frac{L_y(\hat\theta)}{L_y(\theta)} 
\end{equation}
where $L_y(\theta)$ is the likelihood function based on data $y$ and $\hat\theta=\hat\theta(y)$ is the maximum likelihood estimator, a maximizer of the likelihood function, i.e., 
\[ L_y(\hat\theta) = \sup_\theta L_y(\theta). \]
I want to stick with the convention of rejecting $H_0$ when $T_{\theta}(Y)$ is large, so I am using the reciprocal of the usual likelihood ratio statistic.  With this choice of $T_\theta(y)$, setting $p_y(\tilde\theta) = 1$ implies $\prob_{\tilde\theta}\{T_{\tilde\theta}(Y) \geq T_{\tilde\theta}(y)\} = 1$.  This means that $T_{\tilde\theta}(y)$ is at the lower bound of the range of $T_{\tilde\theta}(Y)$ when $Y \sim \prob_{\tilde\theta}$, in other words, $\tilde\theta$ minimizes the function $T_\theta(y)$ with respect to $\theta$, for the given $y$.  By the definition of $T_\theta(y)$ in \eqref{eq:lrt}, we have $T_\theta(y) \geq 1$, and equality is obtained if and only if $\tilde\theta$ maximizes $L_y(\theta)$.  Therefore, the ``maximum p-value estimator'' $\tilde\theta$ is just the maximum likelihood estimator $\hat\theta$.   

There is nothing new here in terms of theory, at most all that changes is how the information in data is summarized in the p-value function \eqref{eq:function} for the goal of inference on $\theta$.  The key point is that the usual tasks associated with statistical inference are conceptually straightforward once the p-value function has been found, and the resulting inference is \emph{valid} in the sense that there are provable guarantees on the frequentist error rates.  This sort of unification of the common inferential tasks should make the concepts easier for students.  Another point is that the properties of the p-value-based procedures discussed above do not require asymptotic justification.  This is interesting from a theoretical point of view, but this also has pedagogical consequences.  In particular, even the students who can follow the technical details of the asymptotic convergence theorems have difficulty seeing how it relates to the problem at hand, so removing or at least down-weighting the importance of asymptotics would be beneficial.  This is not to say that asymptotic considerations are not useful; see Section~\ref{SS:asymptotic}.  

The take-away message is that the p-value function \eqref{eq:function} is a useful and arguably fundamental object for the purpose of statistical inference.  Students will see that evaluating the p-value function is the biggest challenge and, fortunately, this is a concrete mathematical/computational problem with lots of tools available to solve it.  

\subsection{First examples}
\label{SS:examples1}

Here I will present a few of the standard examples from an introductory statistical inference course from the point of view described above, treating the p-value function as the key object.  Now that it is time to put this proposal into action, an obvious question arises: which test statistic $T_\theta(Y)$ to use?  For the sake of having a consistent presentation, along with other reasons discussed in Section~\ref{S:outline}, I will take $T_\theta(Y)$ to be the likelihood ratio statistic in \eqref{eq:lrt}.  Of course, other choices of $T_\theta(Y)$ can be used, e.g., based on well-known pivots for the particular models, but I will leave this decision to the instructor.  

\subsubsection{Normal model}

\begin{figure}[t]
\begin{center}
\subfigure[Normal: $n=10$, $\ybar=7$]{\scalebox{0.6}{\includegraphics{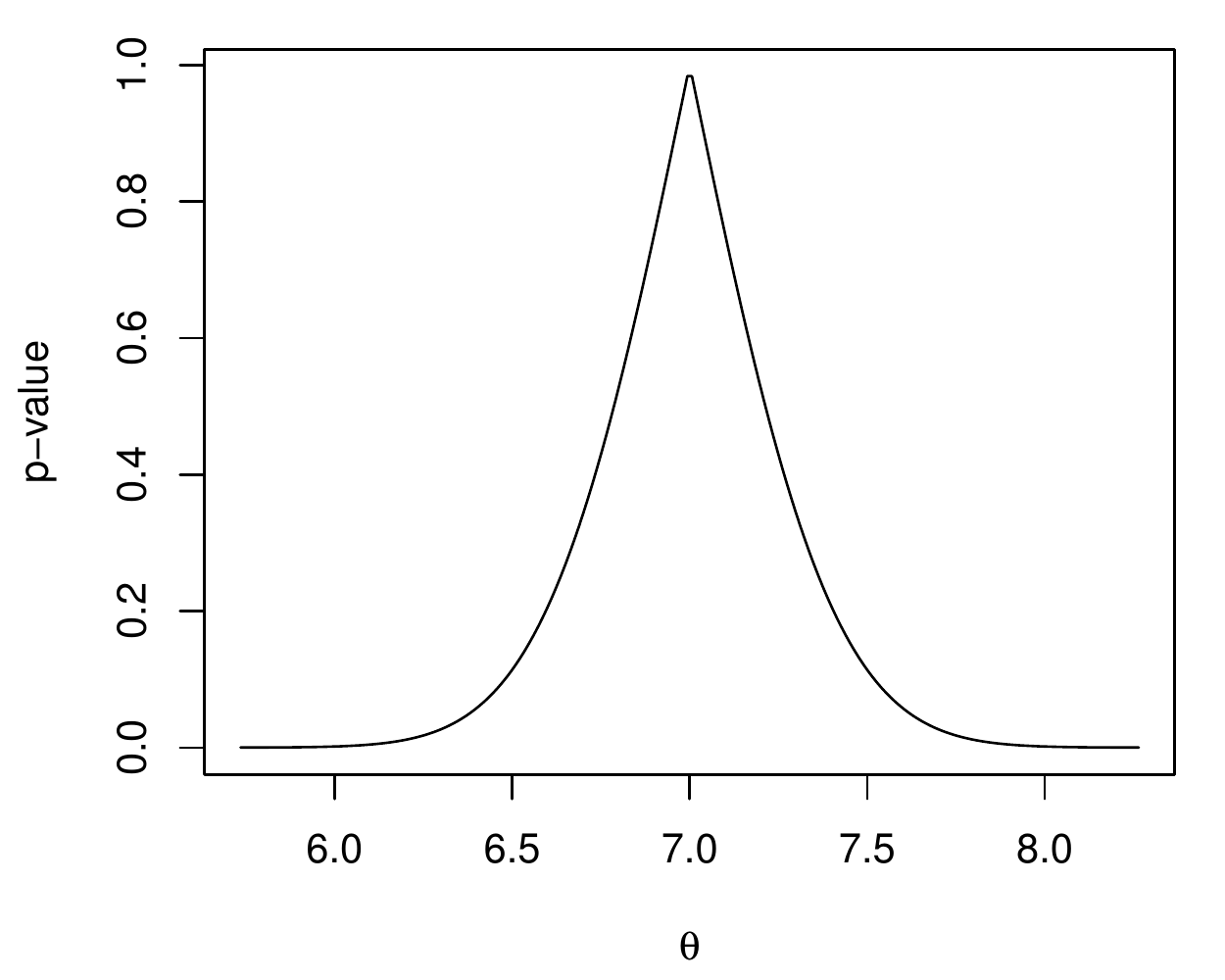}}}
\subfigure[Uniform: $n=10$, $y_{(n)}=7$]{\scalebox{0.6}{\includegraphics{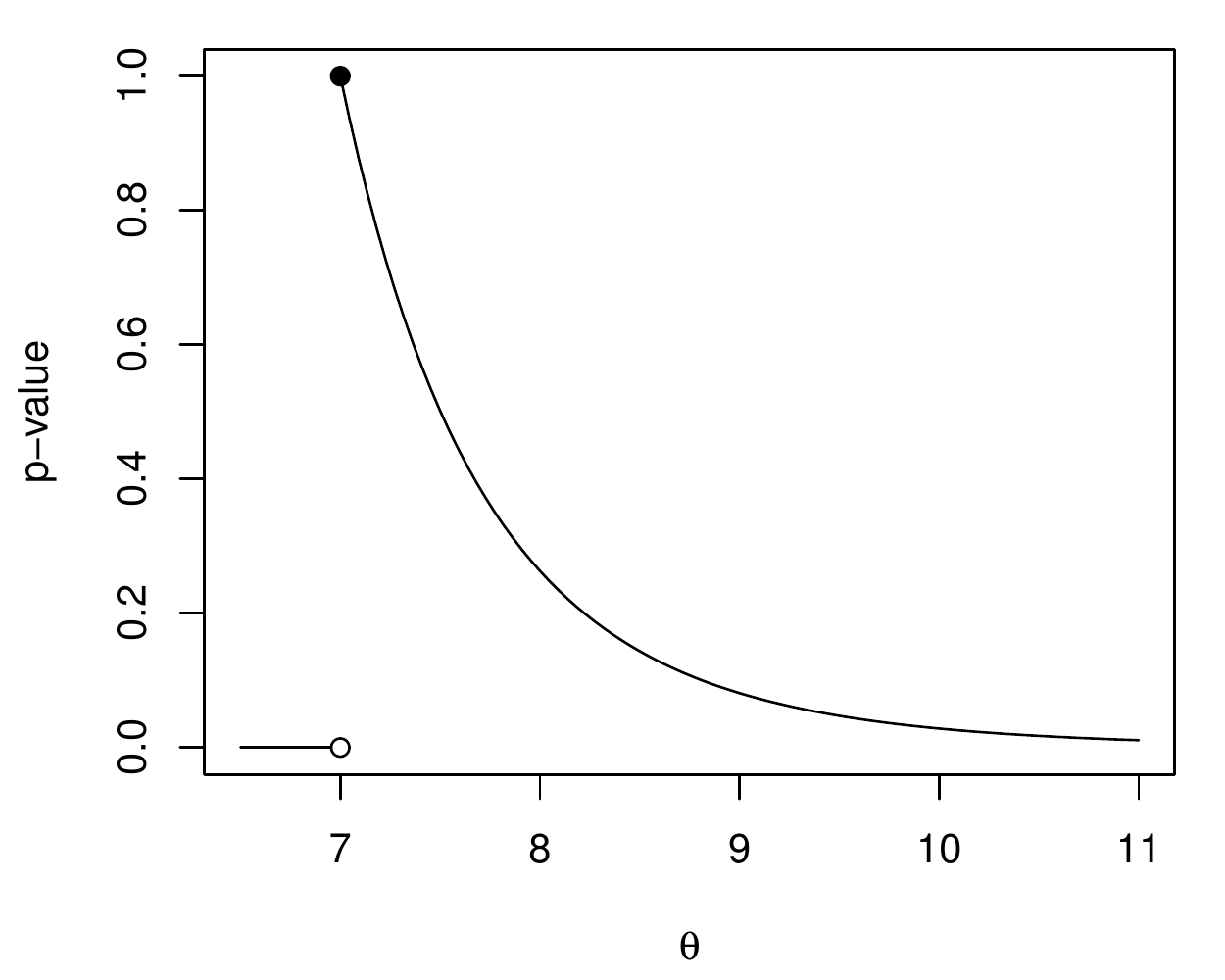}}}
\subfigure[Exponential: $n=10$, $\ybar=7$]{\scalebox{0.6}{\includegraphics{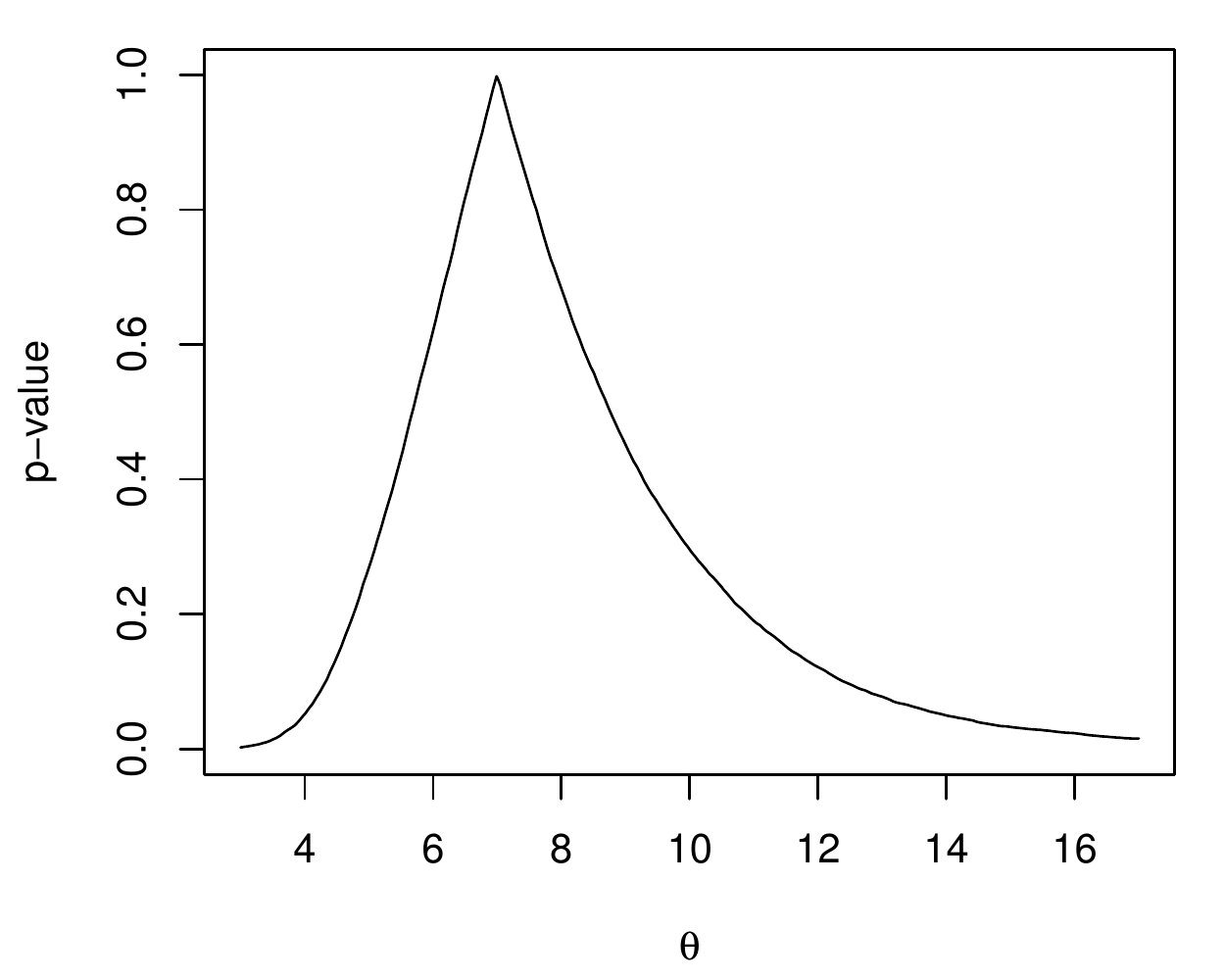}}}
\subfigure[Binomial: $n=20$, $y=13$]{\scalebox{0.6}{\includegraphics{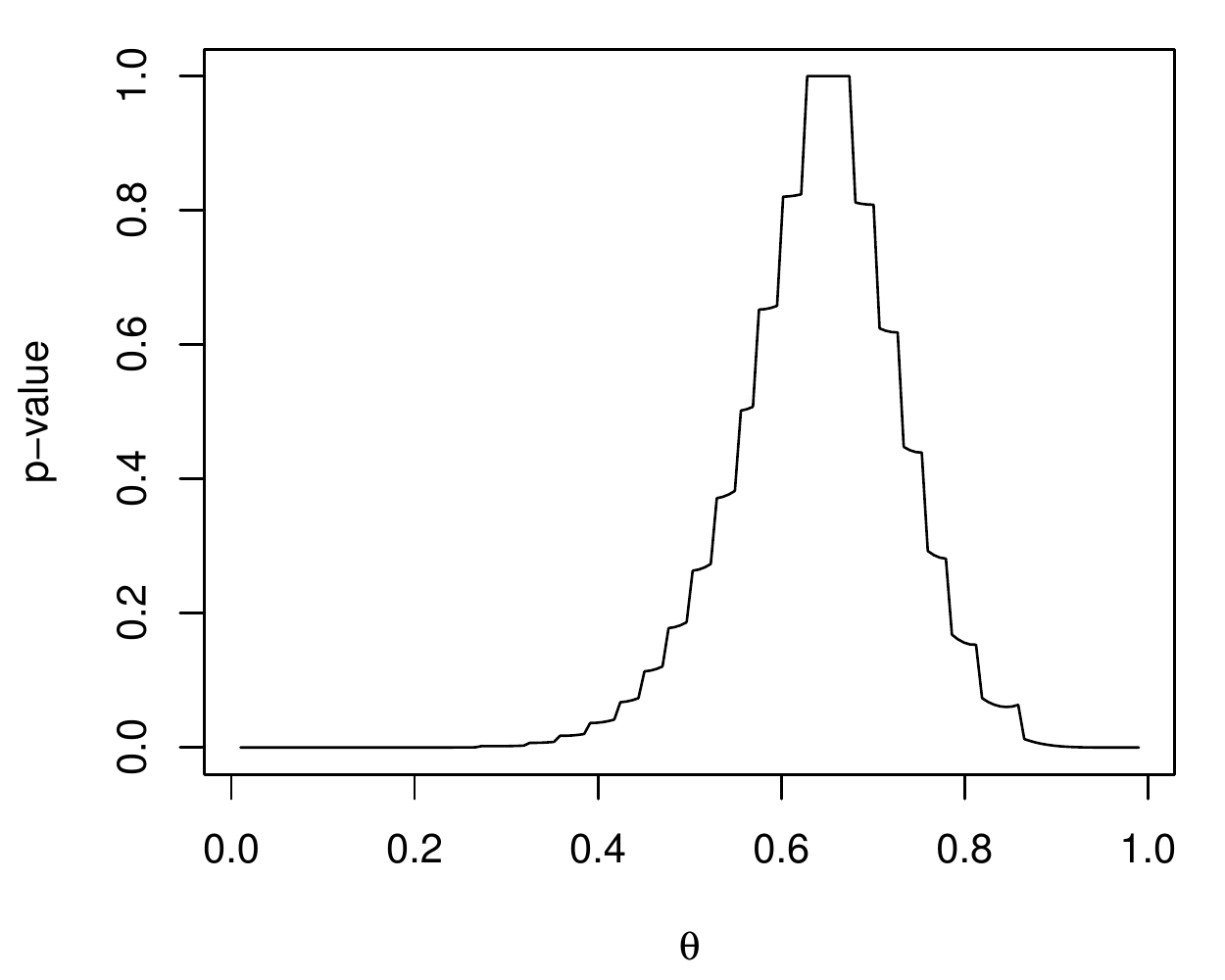}}}
\end{center}
\caption{Plots of the p-value function for the four examples in Section~\ref{SS:examples1}.}
\label{fig:examples1}
\end{figure}

Let $Y=(Y_1,\ldots,Y_n)$ be an independent and identically distributed (iid) sample from a normal distribution $\nm(\theta,1)$ with known variance but unknown mean.  The maximum likelihood estimator is $\hat\theta=\Ybar$, the sample mean, and the likelihood ratio statistic is 
\[ T_{\theta}(Y) = \frac{L_Y(\hat\theta)}{L_Y(\theta)} = e^{\frac{n}{2} (\Ybar - \theta)^2}. \]
It is well known that $2\log T_{\theta}(Y) = n(\Ybar - \theta)^2$ has a $\chisq(1)$ distribution, so the p-value function \eqref{eq:function} is simply 
\[ p_y(\theta) = 1 - G\bigl(2\log T_{\theta}(y)\bigr), \]
where $G$ is the $\chisq(1)$ distribution function.  A plot of this p-value function is shown in Figure~\ref{fig:examples1}(a) for the case of $n=10$ and $\ybar=7$.  It is straightforward to check that the p-value interval \eqref{eq:region} is exactly the standard z-interval found in textbooks.

\subsubsection{Uniform model}
\label{SSS:uniform}

Let $Y=(Y_1,\ldots,Y_n)$ be an iid sample from $\unif(0,\theta)$, a continuous uniform distribution on the interval $(0,\theta)$, where $\theta > 0$ is unknown.  The maximum likelihood estimator, in this case, is $\hat\theta=Y_{(n)}$, the sample maximum.  Using the likelihood ratio statistic, the p-value function \eqref{eq:function} is easily seen to be 
\[ p_y(\theta) = \begin{cases} F_n(y_{(n)} / \theta) & \text{if $\theta \geq y_{(n)}$} \\ 0 & \text{if $\theta < y_{(n)}$}, \end{cases} \]
where $F_n$ is the $\bet(n,1)$ distribution function.  A plot of this p-value function is shown in Figure~\ref{fig:examples1}(b), with $n=10$ and $y_{(n)}=7$.  Note, also, that the equation $p_y(\theta)=\alpha$ has exactly one solution, i.e., $\theta = y_{(n)} / F_n^{-1}(\alpha)$.  Therefore, the exact $100(1-\alpha)$\% p-value confidence interval for $\theta$ is $[ y_{(n)}, y_{(n)} / F_n^{-1}(\alpha) )$.

\subsubsection{Exponential model}

Let $Y=(Y_1,\ldots,Y_n)$ be an iid sample from the exponential distribution with unknown mean $\theta > 0$.  The maximum likelihood estimator is $\hat\theta=\Ybar$, and the likelihood ratio is 
\[ T_{\theta}(Y) = \frac{L_Y(\hat\theta)}{L_Y(\theta)} = \Bigl( \frac{\Ybar}{\theta} \Bigr)^{-n} e^{n(\Ybar / \theta - 1)}. \]
The distribution of $T_{\theta}(Y)$ is not of a standard form, but there are several ways to evaluate the p-value function.  First, level sets of the function $z \mapsto z^{-n} e^{n(z-1)}$ are intervals, and can be found numerically using bisection, say; then the fact that $n\Ybar$ has a $\gam(n,\theta)$ distribution can be used to evaluate the p-value numerically using, e.g., the {\tt pgamma} function in R.  Second, since the distribution of $\Ybar/\theta$ is free of $\theta$, the p-value function can be approximated using Monte Carlo, using only a single Monte Carlo sample; see Section~\ref{SS:monte}.  A plot of the p-value function based on $n=10$ and $\ybar=7$ is shown in Figure~\ref{fig:examples1}(c); observe the asymmetric shape compared to the normal model.  The p-value-based confidence interval in \eqref{eq:region} can be found numerically; in this case, the 95\% confidence interval is $(3.98, 14.07)$.

\subsubsection{Binomial model}

Let $Y \sim \bin(n,\theta)$ be a binomial observation, with the number of trials $n$ known but success probability $\theta \in (0,1)$ unknown.  The maximum likelihood estimator is $\hat\theta = Y/n$, and the likelihood ratio statistic is 
\[ T_{\theta}(Y) = \Bigl( \frac{Y}{n\theta} \Bigr)^Y \Bigl( \frac{n-Y}{n(1-\theta)} \Bigr)^{n-Y}. \]
There is no clean expression for the corresponding p-value but, since the binomial distribution is supported on the finite set $\{0,1,\ldots,n\}$, it is possible to enumerate all the values of $Y$ such that $T_{\theta}(Y) \geq T_{\theta}(y)$, where $y$ is the observed count.  Then the p-value function \eqref{eq:function} can be computed by just summing up the probability masses associated with these values of $Y$.  A plot of this p-value function, based on $n=20$ and $y=13$ is shown in Figure~\ref{fig:examples1}(d).  The stair-step shape of the curve is a consequence of the discreteness.  Note that this p-value-based approach to confidence interval construction is very different from the usual Wald interval that is typically taught, but different is arguably better in this case given that the latter is known to be problematic \citep{bcd2001}.  The numerical results in this case are similar, however: the 95\% p-value interval is $(0.42, 0.86)$ and the corresponding Wald interval is $(0.44, 0.86)$.

\section{Beyond the basics}
\label{S:beyond}

\subsection{Computation}
\label{SS:monte}

Except for basic problems, like those in Section~\ref{SS:examples1}, the p-value function cannot be written in closed-form.  However, it is straightforward to obtain a Monte Carlo approximation thereof.  That is, if $Y^{(1)},\ldots,Y^{(M)}$ are independent samples from the model $\prob_{\theta}$, where $M$ is large, then the law of large numbers implies that 
\begin{equation}
\label{eq:monte}
p_y(\theta) \approx \frac1M \sum_{m=1}^M I_{\{T_{\theta}(Y^{(m)}) \geq T_{\theta}(y)\}}. 
\end{equation}
This can be repeated for as many values of $\theta_0$ as necessary to, say, draw a graph of the p-value function.  Depending on the task at hand, certain properties of the approximation to $p_y(\cdot)$ must be extracted.  For example, in hypothesis testing, based on the general formula \eqref{eq:pvalue.sup}, optimization of the right-hand side of \eqref{eq:monte}, with respect to $\theta$, is needed.  Similarly, to obtain the confidence region \eqref{eq:region}, solutions to the equation $p_y(\theta) = \alpha$ are needed.  There are a variety of ways to solve each of these problems.  A simple naive solution can be obtained by taking a sufficiently fine discretization of the parameter space, e.g., approximating the supremum in $p_y(\Theta_0)$ by the maximum of $p_y(\vartheta)$ for $\vartheta$ ranging over a finite grid spanning $\Theta_0$.  Alternatively, one can apply standard optimization and root-finding procedures to the approximation in \eqref{eq:monte}.  For example, in R, the {\tt uniroot} and {\tt optim} functions can be used for root-finding and optimization.  For numerical stability, it is advised that one use the same seed in the random number generator when evaluating both $p_y(\theta)$ and $p_y(\theta')$.  More details can be found in the examples in Section~\ref{S:examples2}.   

An obvious concern is that the proposed Monte Carlo approximation in \eqref{eq:monte} might be terribly expensive, especially if it needs to be repeated for several values of $\theta$.  As a first idea towards speeding things up, observe that $T_\theta(Y)$ often depends only on some function of $Y$, i.e., a sufficient statistic, so it may not be necessary to simulate copies of the full data $Y$ at each step of the Monte Carlo approximation.  Second, it may be that the problem under consideration has a special structure so that the distribution of $T_\theta(Y)$, under $Y \sim \prob_\theta$, does not depend on $\theta$, i.e., that $T_\theta(Y)$ is a \emph{pivot}.  In that case, the same samples $Y^{(1)},\ldots,Y^{(M)}$ can be used for all values of $\theta$, which significantly speeds up the computation of the p-value function at different parameter values.  Third, in light of the improved speed in the pivotal case, it is natural to ask if it is possible to use only a single Monte Carlo sample even in the non-pivotal case.  At least in some cases, the answer is YES.  In particular, one can employ an \emph{importance sampling} technique \citep[e.g.,][]{lange2010book}, whereby a single Monte Carlo sample $Y^{(1)},\ldots,Y^{(M)}$ is drawn from a distribution with (joint) density function $f(y)$, and \eqref{eq:monte} is replaced by 
\[ p_y(\theta) \approx \frac1M \sum_{m=1}^M I_{\{T_{\theta}(Y^{(m)}) \geq T_{\theta}(y)\}} \frac{p_{\theta}(Y^{(m)})}{f(Y^{(m)})}, \]
where $p_\theta(y)$ is the (joint) density function for $Y \sim \prob_\theta$.  Of course, the quality of this importance sampling approximation depends heavily on the choice of $f$, so this needs to be addressed, but there are some general rules of thumb available.  



\subsection{Handling nuisance parameters}
\label{SS:nuisance}

Standard textbooks do not adequately address the difficulties that arise from the presence of nuisance parameters.  Suppose that the unknown parameter $\theta$ can be partitioned as $(\psi, \lambda)$, where $\psi$ is the interest parameter and $\lambda$ is the nuisance parameter; both $\psi$ and $\lambda$ can be vectors.  Except for a bit about profile likelihood, as I discuss below, and perhaps a few normal examples where marginalization is relatively easy, textbooks focus primarily on asymptotics and Wald-style methods where an estimator $\hat\lambda$ is plugged in for $\lambda$ in the asymptotic variance of the estimator $\hat\psi$ of $\psi$.  The simplicity of this approach comes at a price: the plug-in estimator of the variance can over- or under-estimate the actual variance, so it may not adequately address uncertainty.  Marginalization is one of the most difficult problems in statistical inference, so there is no way to give a completely satisfactory solution in a first course on the subject.  However, there are some general and relatively simple techniques that can be presented to students which, together with a warning about the difficulty of marginal inference on $\psi$, ought to suffice.  

Here I will present two distinct approaches to marginalization, both relying on optimization.  The first is a familiar one, namely, \emph{profiling}.  In particular, the profile likelihood ratio statistic is 
\begin{equation}
\label{eq:plrt}
T_{\psi}(Y) = \frac{\sup_{\psi,\lambda} L_Y(\psi, \lambda)}{\sup_\lambda L_Y(\psi, \lambda)}. 
\end{equation}
The right-hand side does not explicitly depend on the nuisance parameter $\lambda$, but its distribution might.  There are special cases where the distribution of the profile likelihood ratio $T_\psi(Y)$ is free of $\lambda$, in which case a ``marginal p-value function'' can be obtained without knowing $\lambda$, even if Monte Carlo methods are needed.  Checking that the distribution of the profile likelihood ratio does not depend on the nuisance parameter might be a difficult exercise, but there are cases where it can be done; see, Section~\ref{S:examples2}.  Problem-specific considerations might also lead to a different choice of test statistic, other than profile likelihood ratio, that has a $\lambda$-free distribution.  There are also some reasonable $\lambda$-free approximations available, as discussed in Section~\ref{SS:asymptotic}.  

The second optimization-based approach to marginalization starts with the formula \eqref{eq:pvalue.sup} for the p-value under a composite null hypothesis.  Indeed, the marginal inference problem can be reduced to one that involves a composite null hypothesis, where the null specifies no constraints on the nuisance parameter.  This suggests that a marginal p-value function for $\psi$ can be expressed, with a slight abuse of notation, as follows:
\[ p_y(\psi) = \sup_\lambda p_y(\psi, \lambda), \]
where the right-hand side is the largest of the original p-values in \eqref{eq:pvalue} corresponding to a fixed value of the interest parameter.  So, those points about optimization of the p-value function discussed in Section~\ref{SS:monte} are relevant again here for marginal inference.

\subsection{Asymptotic approximations}
\label{SS:asymptotic}

An interesting feature of the proposed approach is that one could potentially fill a first course on statistical inference without any serious discussion of asymptotic theory.  I do not necessarily recommend that asymptotic theory be left out entirely, but I think its importance needs to be downplayed compared to the traditional first course.  Students should be encouraged to do exact (analytical or numerical) calculations whenever possible, only appealing to approximations when the exact calculations cannot be done, either because the computations are too hard or because the model assumptions are too vague to determine what exact calculations need to be done.  In my opinion, it is in this sense that the relevant asymptotic theory should be presented in a first statistical theory course.  

To be concrete, let $Y=(Y_1,\ldots,Y_n)$ be an iid sample from a distribution $\prob_\theta$, with $\theta$ a scalar, and suppose that the likelihood ratio statistic \eqref{eq:lrt} is used to determine the p-value; the comments to be made here apply almost word-for-word to the profile likelihood ratio statistic \eqref{eq:plrt} for marginal inference.  Perhaps the most important asymptotic result in a first statistical theory course is the theorem of \citet{wilks1938}, which gives a large-sample approximation to the distribution of the likelihood ratio statistic, i.e., under suitable regularity conditions, 
\[ 2 \log T_{\theta}(Y) = 2 \log\frac{L_Y(\hat\theta)}{L_Y(\theta)} \to \chisq(1) \quad \text{in distribution}. \]
Therefore, if the regularity conditions hold, then the p-value can be approximated by 
\[ p_y(\theta) \approx 1 - G\bigl(2\log T_{\theta}(y) \bigr), \]
where $G$ is the $\chisq(1)$ distribution function.  In this case, all the relevant calculations for hypothesis testing and/or interval estimation are straightforward.  More refined higher-order approximation results are available \citep[e.g.,][]{brazzale.davison.reid.2007}, but these may be too advanced for a first course.   

My position on the role of asymptotic theory might be controversial, so let me elaborate a bit here in closing.  Students who will choose to get more advanced training will learn more about asymptotic theory which, e.g., can be used to justify a choice of $T_\theta(Y)$.  But the main goal of this first statistics theory course should be that students develop a basic understanding of what statistical inference is and how it can be done; to me, making clear that the primary role played by asymptotic theory is for simple approximations is a necessary step toward this goal.

\section{More challenging examples}
\label{S:examples2}

\subsection{Shifted exponential model}
\label{SS:shexp}

\begin{figure}[t]
\begin{center}
\subfigure[Shifted exponential, Sec.~\ref{SS:shexp}]{\scalebox{0.6}{\includegraphics{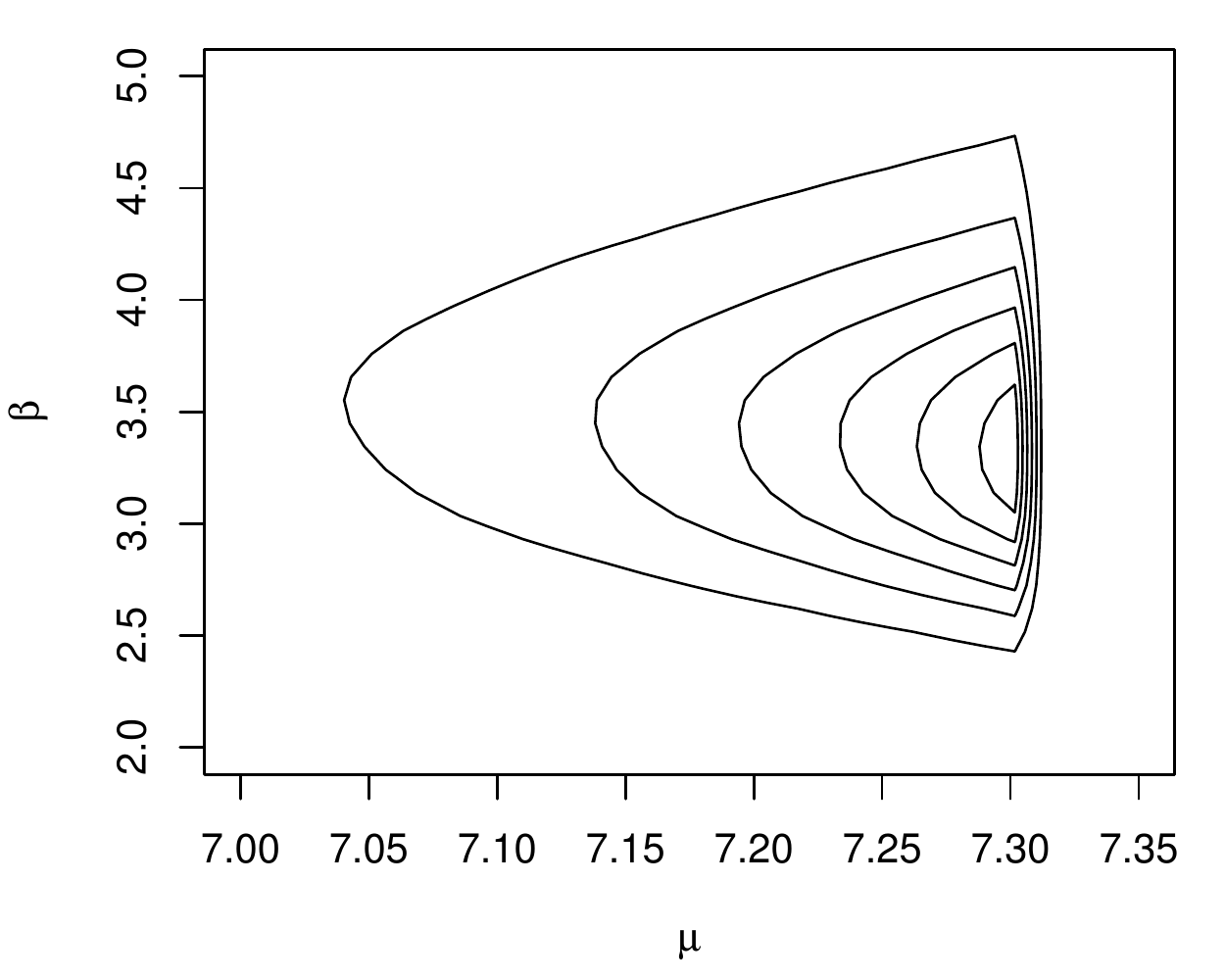}}}
\subfigure[Normal random-effects, Sec.~\ref{SS:random.effects}]{\scalebox{0.6}{\includegraphics{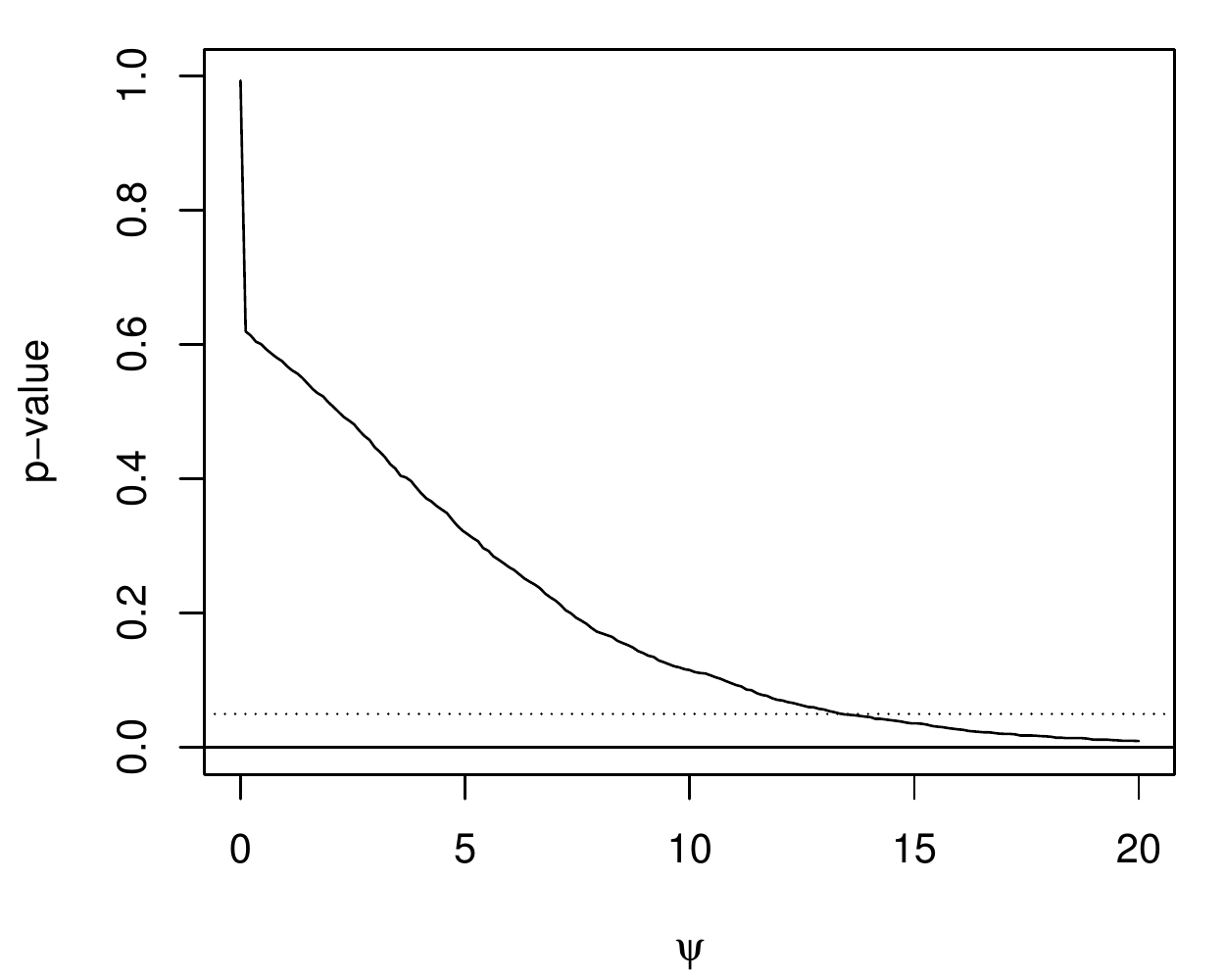}}}
\subfigure[Bivariate normal, Sec.~\ref{SS:bvn}]{\scalebox{0.6}{\includegraphics{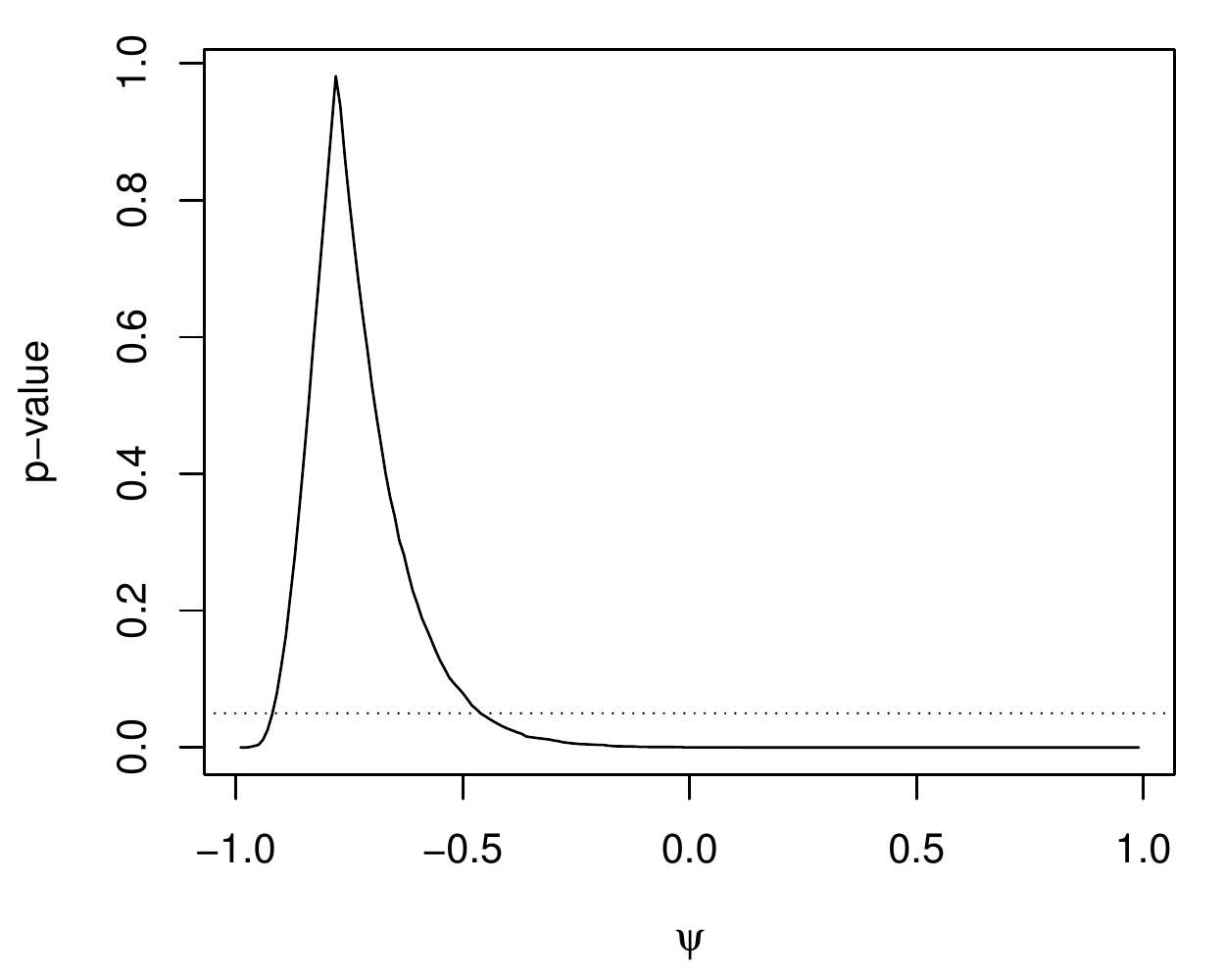}}}
\end{center}
\caption{Plots of the p-value function for the three examples in Section~\ref{S:examples2}.}
\label{fig:examples2}
\end{figure}

Let $Y=(Y_1,\ldots,Y_n)$ be an iid sample from a shifted exponential distribution with common density function $y \mapsto \beta^{-1} e^{-(y - \mu)/\beta}$, for $y \geq \mu$, where $\theta=(\mu,\beta)$ is unknown, where $\mu$ is a location parameter and $\beta$ is a scale parameter.  This is a special case of class of non-regular problems considered in \citet{smith1985}, known to be relatively difficult since the usual asymptotic theory for, say, the maximum likelihood estimator does not hold.  In this case, the profile likelihood ratio is 
\[ T_{\theta}(Y) = \begin{cases} 
\bigl\{\frac{1}{\beta} \sum_{i=1}^n (Y_i - Y_{(1)}) \bigr\}^{-n} e^{\frac{1}{\beta} \sum_{i=1}^n (Y_i - \mu) - n}, & \text{if $Y_{(1)} \geq \mu$}, \\
\infty, & \text{if $Y_{(1)} < \mu$}.
\end{cases}
\]
From here, it is relatively easy to see that $T_{\theta}(Y)$ is a pivot, i.e., if $\theta$ is the true value of the parameter, the distribution of $T_\theta(Y)$ does not depend on $\theta$.  This makes evaluation of the p-value function, via Monte Carlo, straightforward.  For simulated data of size $n=25$ with true values $\mu=7$ and $\beta=3$, a plot of the (bivariate) p-value function is shown in Figure~\ref{fig:examples2}(a).  Note the non-elliptical shape, indicative of a ``non-regular'' problem.

\subsection{Normal random-effects model}
\label{SS:random.effects}

A simple normal random-effects model assumes that $Y=(Y_1,\ldots,Y_n)$ are independently distributed, with $Y_i \sim \nm(\mu_i, \sigma_i^2)$, $i=1,\ldots,n$, where the means $\mu_1,\ldots,\mu_n$ are unknown, but the variances $\sigma_1^2,\ldots,\sigma_n^2$ are taken to be known.  The ``random-effects'' portion of the model comes from the assumption that $\mu_1,\ldots,\mu_n$ are iid $\nm(\lambda,\psi^2)$ samples, where $\theta=(\psi,\lambda)$ is unknown.  Here $\psi \geq 0$ is the parameter of interest.  

This model can be recast in a non-hierarchical form; that is, $Y_1,\ldots,Y_n$ are independent, with $Y_i \sim \nm(\lambda, \sigma_i^2 + \psi^2)$, $i=1,\ldots,n$.  Here the maximizer of the likelihood over $\lambda$, for given $\psi$, is $\hat\lambda_\psi = \sum_{i=1}^n w_i(\psi) Y_i / \sum_{i=1}^n w_i(\psi)$, where $w_i(\psi) = 1/(\sigma_i^2 + \psi^2)$, $i=1,\ldots,n$.  From here 
it is easy to write down the profile likelihood,
\[ L_Y(\psi, \hat\lambda_\psi) = \prod_{i=1}^n (\sigma_i^2 + \psi^2)^{-1/2} e^{-\frac{1}{2(\sigma_i^2 + \psi^2)} (Y_i - \hat\lambda_\psi)^2}, \]
and the profile likelihood ratio $T_\psi(Y)$, as in \eqref{eq:plrt}, can be evaluated numerically using an optimization routine.  Moreover, since $\lambda$ is a location parameter, one can see that the distribution of $T_\psi(Y)$ does not depend on $\lambda$.  Therefore, any choice of $\lambda$ (e.g., $\lambda=0$) will suffice for computing the p-value function for $\psi$ based on Monte Carlo.      

For a concrete example, consider the SAT coaching problem presented in \citet{rubin1981}.  Here $n=8$ coaching programs are evaluated, and inference on $\psi$ is required.  In particular, the case of $\psi=0$ is of inferential importance, as it indicates that there is no difference between the various coaching programs.  Figure~\ref{fig:examples2}(b) shows a plot of the marginal p-value function for $\psi$ for the given data, and the corresponding 95\% confidence interval is $[0, 13.40)$.  Since the interval contains zero, one cannot exclude the possibility that the coaching programs have no effect, consistent with Rubin's conclusion.  Note also that the confidence interval in this case has guaranteed frequentist coverage properties, while other approaches, based only on asymptotics might not be justifiable here, since only $n=8$ samples are available.  


\subsection{Bivariate normal model}
\label{SS:bvn}

Consider a sample of independent observations $Y=\{(Y_{i1},Y_{i2}): i=1,\ldots,n\}$ from a bivariate normal distribution where the two means, two variances, and correlation are all unknown.  That is, the unknown parameter is $\theta = (\psi,\lambda)$, where the correlation coefficient $\psi$ is the parameter of interest, and $\lambda = (\mu_1,\mu_2,\sigma_1,\sigma_2)$ is the nuisance parameter.  From the calculations in \citet{sun.wong.2007}, the profile likelihood ratio is 
\[ T_{\psi}(Y) = \bigl\{ (1-\psi\hat\psi) \, / \, (1-\psi^2)^{1/2}(1-\hat\psi^2)^{1/2} \bigr\}^n, \]
where $\hat\psi$ is the sample correlation coefficient.  A well known property of the correlation coefficient is that it does not change if data are subjected to a linear transformation.  In this case, this implies that the distribution of $T_\psi(Y)$ does not depend on $\lambda$.  Therefore, the p-value function for $\psi$ can be evaluated via Monte Carlo, by simulating from a bivariate normal with any convenient choice of $\lambda$.   

For an illustration, I revisit the example in \citet{sun.wong.2007}.  The data, from  \citet{levine1999}, measure the increase in energy use $y_1$ and the fat gain $y_2$ for $n=16$ individuals, and the sample correlation coefficient is $\hat\psi = -0.77$.  A plot of the p-value function for $\psi$ is shown in Figure~\ref{fig:examples2}(c), based on Monte Carlo.  The corresponding 95\% confidence interval is $(-0.918, -0.461)$, which is very similar in this case to Fisher's classical interval based on the distribution of $z=\frac12 \log\{(1+\hat\psi)/(1-\hat\psi)\}$.  





\section{On implementing the proposal}
\label{S:outline}

Here I will describe how I would teach a course based on this proposal.  First, the usual prerequisite for an introductory statistical inference course is a semester of calculus-based probability, in which students would have learned various things, including the definitions and properties of the standard distributions.  So, except for possibly giving a brief review at the beginning of the course, I would not cover probability topics specifically.  

I have advocated here a general approach based on likelihood or profile likelihood ratios, and I have two reasons for doing this.  First, having a sort of fixed choice for the test statistic can give the presentation some needed unification, compared to using the ``best'' or ``most convenient'' choice for each problem.  Second, by putting an emphasis on likelihood, some of the familiar topics from a standard introductory statistical inference course have a natural place in this new type of course.  In particular:
\begin{itemize}
\item The interpretation of the likelihood function as providing a ``ranking'' of the parameter values in terms of how well the corresponding model fits the given data is important, motivating  maximum likelihood estimation and also the comparison of $L_Y(\theta)$ to $L_Y(\hat\theta)$ in this proposed approach.  These discussions about likelihood also help to make clear to students that a change of perspective is needed to go from thinking about sampling models for data to thinking about inference based on observed data.  
\vspace{-2mm}
\item The notion of sufficient statistics is fundamental, and can be presented here, via the factorization theorem, as the function of data upon which the likelihood ratio depends.  Sufficiency is helpful in the present approach mainly because it can be used to simplify evaluation of the p-value function.   
\vspace{-2mm}
\item The quadratic approximation of the log-likelihood function is key to all the relevant (likelihood-based) asymptotic results presented in a first statistical inference course, so if the new style of course also focuses on likelihood ratios, these results can be seamlessly included based on the discussion in Section~\ref{SS:asymptotic}.  
\end{itemize}

After introducing likelihood ratios and other relevant background, the course can now proceed to  inference based on p-values.  I would begin by presenting, in an informal way, a very basic hypothesis testing problem to motivate the p-value.  From here, I would follow with the formal definition of the p-value function and a detailed demonstration of the properties it satisfies, as discussed in Section~\ref{SS:main}.  Then I would proceed to work out some relatively simple examples, such as those presented in Section~\ref{SS:examples1}.  Various results are used to solve these examples, e.g., that $Y_{(n)}/\theta$ in the $\unif(0,\theta)$ problem of Section~\ref{SSS:uniform} has a beta distribution, and these could be assigned as homework.   

The next part of the course, based on the ideas presented in Section~\ref{S:beyond}, is where things start to get more interesting and new.  Here students will be introduced to some basic computational tools needed to implement the proposed approach for statistical inference based on the p-value.  Depending on the background of students in the class, the instructor may need to take some time to introduce a statistical software package, and I would recommend using R.  This is time well-spent, I believe, because students will need this background anyway and, moreover, students need to understand that no serious work can be done without knowledge of both theory and computation.  In the discussion of the basic Monte Carlo strategy, I would highlight the importance of pivots; this concept appears in standard textbooks but is not given the emphasis it deserves.  In this context, pivots are specifically helpful for simplifying and accelerating the Monte Carlo approximations.  Marginalization, as discussed in Section~\ref{SS:nuisance}, is a difficult problem that requires care, and both distributional and computational tricks can be employed for this purpose.  Then, finally, asymptotic theory can be presented as a means to get a good approximation to the p-value function in complicated problems.  Of course, the approximation theorem, with the required regularity conditions, should be carefully stated and maybe even proved.  Working numerical examples can be discussed along the way in class to compare the results of the various approaches: exact analytical, Monte Carlo-based, and asymptotically approximate solutions.  

The course would end with a discussion of several non-trivial examples implementing the various techniques, perhaps with a comparison with other methods.  Depending on time, I would also discuss briefly what other things students would learn in, say, a more advanced course.  This includes the ``optimal'' choice of $T_\theta(Y)$ and how to deal with both theory and computations when $\theta$ is high-dimensional.  

Readers may notice that my proposed course leaves out some other topics that may occasionally be covered in this first statistical inference course, such as Neyman--Pearson optimality, minimum variance unbiased estimation (including the Cram\'er--Rao inequality, completeness, and the Rao--Blackwell and Lehmann--Sch\'effe theorems), Bayesian inference, etc.  These are, indeed, important topics but I consider them to be relevant only to students who will choose to specialize in statistics.  So, for a first statistics theory course, whose audience will likely include as many students who ultimately will not specialize in statistics, it is best to leave these more advanced topics out.  As a specific example, consider the sort of Bayesian inference that is typically included in such a course.  Textbooks focus primarily on deriving Bayes estimators under conjugate priors, which is not representative of modern Bayesian analysis.  Giving a very brief and out-dated presentation of a relatively advanced topic is potentially misleading to students and, more importantly, potentially harmful to the subject itself.  In fact, the p-value-centered approach proposed here might actually help students to better understand and appreciate a Bayesian approach.  Bayesian methods require care in choice of prior and often require Monte Carlo methods to compute the posterior.  To many students, this  Bayesian approach appears to be ``harder'' than the classical one based on simple asymptotic approximations.  If students see that a valid non-Bayesian approach also requires care in the setup and Monte Carlo methods to compute the p-value function, then they can make a meaningful and less superficial comparison between a Bayesian and non-Bayesian approach.

\section{Discussion}
\label{S:discuss}

In this paper, I have proposed an alternative approach to teaching the first statistical inference course to senior undergraduates or beginning graduate students with a calculus-based probability background.  The basic idea is that the p-value function contains relevant information for all tasks related to statistical inference.  Besides the uniform presentation, the resulting inference is valid in the sense that there are provable controls on the frequentist error rates, compared to the classical procedures presented in such courses which, in many cases, are only asymptotically valid.  The price that is paid for these desirable features is that, outside the standard textbook problems, the solutions may not be so simple to write down.  Specifically, the p-value-based solution for most problems will involve numerical methods, including Monte Carlo.  Trading simple analytic solutions with only asymptotic validity for less simple numerical solutions with guaranteed validity seems beneficial to me, so it makes sense to do this in the first statistical inference course.  Indeed, inclusion of numerical methods into a statistical theory course is of broad interest and value, and the proposed course provides an idea for accomplishing this.  

There are some potential downsides to changing the way the first statistical theory course is taught.  One in particular, raised by a referee, is that students might be better served by exposing them to the concepts and vocabulary common among practicing statisticians.  I think it is safe to say that there are serious concerns these days about how statistical methods and reasoning are being used in practice, so perhaps a change is needed.  This proposed course, I think, is a step in the right direction.    

Finally, I want to briefly mention that the proposed approach is not just a simple strategy suitable for teaching in a first statistics theory course---it can be used to solve real problems.  The only obstacle in applying the proposed approach to modern statistical problems is computation; that is, the naive Monte Carlo approximation in \eqref{eq:monte} might be too crude for problems involving moderate- to high-dimensional $\theta$.  Therefore, work is needed to develop efficient Monte Carlo methods for these problems.  So, the computational challenges to implement the proposed approach is not a shortcoming, it is an opportunity for new research and developments.  I believe that the standards of asymptotically valid inference are too low, and I would encourage others to consider raising both their teaching and research above and beyond these norms.

\section*{Acknowledgement}

The author thanks Professor Samad Hedayat as well as the Editor, Associate Editor, and referees for their valuable comments on a previous version of this manuscript.  


\bibliographystyle{apalike}
\bibliography{/Users/rgmartin/Dropbox/Research/mybib}

\end{document}